\documentclass[aps,prl,reprint]{revtex4-1}
\usepackage{blindtext}
\usepackage{epsfig}
\usepackage{color}
\usepackage{amsmath}
\usepackage{amsfonts}
\usepackage{amssymb}
\usepackage{graphicx}%
\usepackage{url}
\usepackage[breaklinks]{hyperref}
\usepackage[english]{babel}
\usepackage{esvect}
\usepackage[normalem]{ulem}
\usepackage[inline]{enumitem}

\hypersetup{colorlinks,citecolor=blue,filecolor=blue,linkcolor=blue,urlcolor=blue}

\usepackage{pdfpages} 

\newcommand{\ket}[1]{\vert #1 \rangle}

\begin{document}
\title{Ramsey Interferometry of Particle-Hole Pairs in Tunnel Junctions}

\author{Tal Goren$^{1,2}$, Karyn Le Hur$^{2}$ and Eric Akkermans$^{1}$ }
\affiliation{$^1$Department of Physics, Technion Israel Institute of Technology, Haifa 32000, Israel}
\affiliation{ $^2$Centre de Physique Th\'{e}orique, \'{E}cole Polytechnique, CNRS, Universit\'{e} Paris-Saclay,  91128 Palaiseau C\'{e}dex, France}

\begin{abstract}
We present a method to probe real-time dynamics in quantum mesoscopic systems using Ramsey interferometry.  
This allows us to explore the effect of interactions on quasi-particles in the time domain. We investigate the dephasing effects of an ohmic environment on an electron-hole pair in a tunnel junction. We show that dynamical Coulomb blockade phenomena can be observed for resistances much smaller than the quantum of resistance. Moreover, the crossover between high and low impedance limits can be probed for a constant resistance by a proper control of the voltage modulation. 
\end{abstract}

\date{\today}
\maketitle

Ramsey experiments are commonly used in systems with discrete energy spectrum to perform accurate spectroscopy of the system \cite{Cohen-Tannoudji_book,Vanier2005,Vion2002,Mooij2003,Chuang2005,atala2013}.  
We present the first  proposal to observe Ramsey interferences in mesoscopic fermionic systems and discuss an application to the time-interferometry of particle-hole pairs in a dissipative tunnel junction.
Ramsey effect was generalized for a system with a continuum set of states for the Schwinger electron-positron pair production from the vacuum \cite{Akkermans2012}.
Ramsey interferometry has also been studied for spin ensembles \cite{Demler2013} and for quasi-condensates in ultra-cold atoms \cite{Kitagawa2010}.

Ramsey interference shows up in the probability to create quasi-particles in the system by a proper time-dependent voltage modulation.
We present a method to relate this probability to a current noise measurement. 
Moreover, we show that the dephasing time of the quasi-particles can be directly read off the Ramsey interference pattern and the time dependence of the dephasing processes can be accessed.

Observing quantum interference effects in mesoscopic systems is a challenging task \cite{Akkermans2007,ImryBook,NazarovBlanterBook,BlanterButtiker2000}.
Mach-Zehnder interferometry has been implemented in one-dimensional systems to measure the coherence length \cite{Roulleau2008,Haack2011} and study dephasing mechanisms \cite{neder2007,LeHur2005,Seelig2001,Gefen2007,Litvin2007,Levkivskyi2008}.
In contrast, we propose time-interferometry which requires only one scatterer (tunnel junction).
The effect of a periodic voltage modulation on the current noise was explored theoretically in Refs. \cite{Pedersen1998,Lesovik1994,Levitov1996} and realized experimentally in Refs. \cite{Gabelli2013,dubois2013}.

In this paper, we consider, as a working example, a tunnel junction (TJ) composed of two metals separated by a thin insulating barrier. But the present scheme can be easily transposed to other situations. The barrier is thin enough to allow electronic transport  by tunnelling. The electrons in the two metals are described by the Hamiltonian 
\begin{equation} \label{H0}
H_{0}=\sum_{k}\epsilon_{Lk}(t)c_{Lk}^{\dagger}c_{Lk}+\sum_{q}\epsilon_{Rq}c_{Rq}^{\dagger}c_{Rq}
\end{equation}
 where $c_{Lk}^{\dagger} \left(c_{Rq}^{\dagger}\right)$  creates an electron in the left (right) lead. 
  The voltage is included in the energies of the left lead
$\epsilon_{Lk}\left(t\right)=\epsilon_{Lk}-eV_{dc}+eV\left(t\right)$. 
$V_{dc}$ refers to a constant bias voltage applied across the TJ and the Ramsey modulations is taken into account though $V(t)$.

Tunneling is described by the Hamiltonian \cite{Bardeen1961,Rogovin1974}
\begin{equation} \label{eq:Tunn_H}
H_{T}=\sum_{kq}\left[T_{kq}c_{Rq}^{\dagger}c_{Lk}+T_{kq}^{*}c_{Lk}^{\dagger}c_{Rq}\right]
\end{equation}
where $T_{kq}c_{Rq}^{\dagger}c_{Lk}$  transfers an electron from the left lead to the right lead, leaving a hole in the left lead.
Such a tunnelling event creates a electron-hole ($e$-$h$) pair in the TJ with energy $\Delta\epsilon=\epsilon_{Lk}-\epsilon_{Rq}-eV_{dc}$. 
Here, the electron and  the hole are produced in different Fermi reservoirs.

Our purpose is to exemplify that a Ramsey protocol can be designed for such many-particle systems. It consists in applying to the TJ a series of two  consequent voltage pulses of width $\tau$ separated by a time $t_0$ such that $t_0\gg\tau$:
\begin{equation}
V(t)=V_0(t)+V_0(t-t_0).
\end{equation}
During the voltage pulses an electron can tunnel from one lead to the other while absorbing an energy $\Delta\epsilon$ from the time-dependent voltage pulse. Note that $e$-$h$ pairs created by a $dc$ voltage have $\Delta\epsilon=0$. 
The response to the voltage pulses is provided by $e$-$h$ pairs  having $\Delta\epsilon\neq0$.

\begin{figure}
\begin{center}
\includegraphics[width=0.5\textwidth]{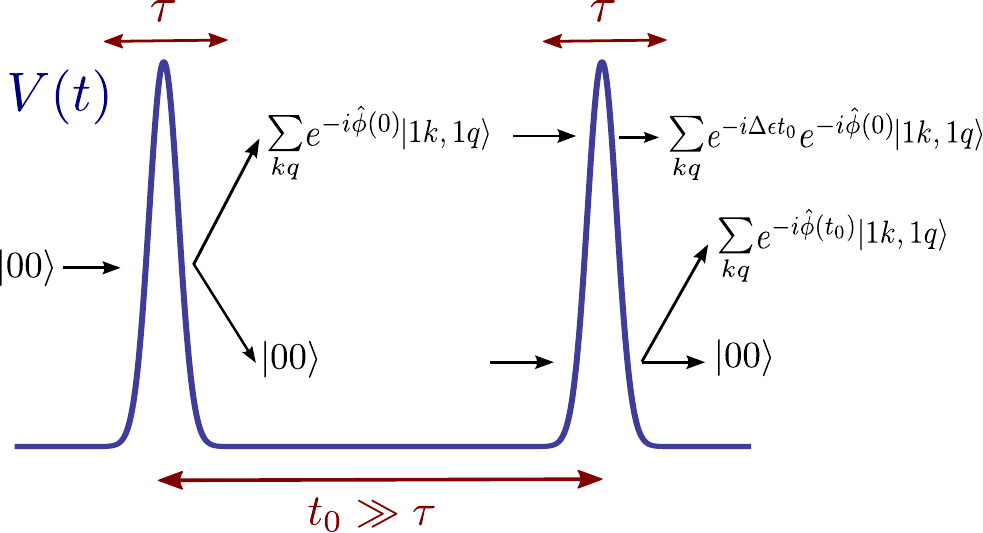}
\end{center}
\vskip -0.5cm
\protect\caption[Ramsey illustration]{ (color online) \it 
Illustration of the  Ramsey protocol describing the probability of an $e$-$h$ pair creation ($\ket{1_k,1_q}$). Two identical voltage pulses of width $\tau$ are applied to the tunnel junction separated by the time delay $t_0$. A $e$-$h$ pair with energy $\Delta\epsilon$ can be created either in the first or in the second pulse. The resulting probability will be the interference between these two quantum amplitudes. In the absence of interactions (or environmental noise) $\hat{\phi}=0$.
\label{fig:Ramsey_illustration}
}
\end{figure}

\begin{figure}
\begin{center}
\includegraphics[width=0.5\textwidth]{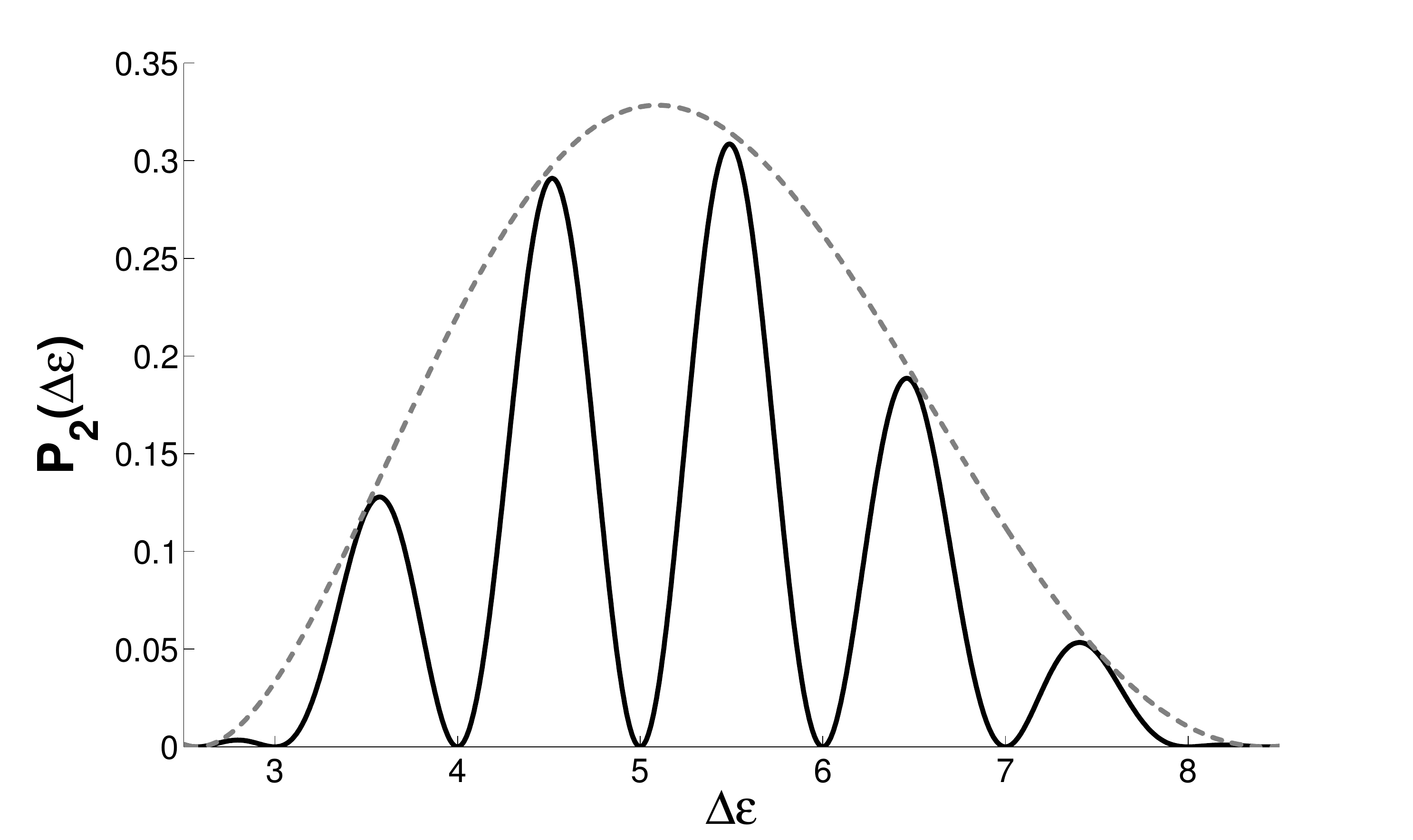}
\end{center}
\protect\caption[Ramsey interference in the  $e$-$h$ creation probability in a tunnel junction]{ \it Probability $P_2(\Delta\epsilon)$ for the creation of   a $e$-$h$ pair for two Gaussian voltage pulses (solid line). The envelope function (dashed line) is $4P_1(\Delta\epsilon)$, where the probability $P_1$  corresponds to a single voltage pulse.  The  time difference between the pulses is $t_0=20\tau$ ($\tau$ is the Gaussian width of the pulses). The energy $\Delta\epsilon$  of the pair is given in units of $2\pi\hbar\over t_0$ and  the single pulse area was chosen to be $\varphi_0=7\pi$, so that the probability vanishes at integer values. The tunnelling matrix element was chosen to be $T=1$ in units of $2\pi\hbar\over t_0$. 
\label{fig:Ramsey_in_TJ}
}
\end{figure}

To lowest order in $T$, the tunnelling matrix element, there are two routes to excite an $e$-$h$ pair, either during the first or the second voltage pulse (see fig. \ref{fig:Ramsey_illustration}). The probability amplitude for the creation of a pair with energy $\Delta\epsilon$ is the coherent sum of the amplitudes of these two processes.
The probability amplitude to create the $e$-$h$ pair during the first pulse is $A_1(\Delta\epsilon,\tau)$. During the free evolution between the pulses it gains a phase $e^{-\frac{i}{\hbar}\Delta\epsilon t_0}$ so that after the two pulses, the probability amplitude for the first path is $A_1(\Delta\epsilon,\tau)e^{-\frac{i}{\hbar}\Delta\epsilon t_0}$.
In the second path, the $e$-$h$ pair is created in the second pulse, and its probability amplitude is  $A_1(\Delta\epsilon,\tau)$. Thus, the probability to create a $e$-$h$ pair with energy $\Delta\epsilon\neq 0$ after the two voltage pulses displays an interference part between the two quantum amplitudes, 
\begin{equation} \label{P2}
P_{2}(\Delta\epsilon,t_0)=P_{1}(\Delta\epsilon,\tau)\left[2+2\cos\left(\Delta\epsilon t_{0}/\hbar \pm\varphi_{0} \right)\right]
\end{equation}
where  $P_{1}=\left|A_1\right|^2$ is the creation probability after a single pulse and $\varphi_0$ is  the pulse area $\varphi_0\equiv\frac{e}{\hbar}\int_0^{\tau} V_0(t)dt$. The single pulse shape determines only the envelope of the Ramsey oscillations. The $\pm$  sign accounts for tunneling processes from the left/right lead to the right/left lead. Eq. \ref{P2} demonstrates that the  Ramsey effect is a manifestation of the interference between two quantum amplitudes just as in Young's two-slit experiment, except that in the Ramsey setup the phase is accumulated over time instead of space. 
The first term in Eq. \eqref{P2} is the classical sum of the probabilities and the second term is the quantum interference which oscillates with the phase difference $\Delta\epsilon \,  t_0 / \hbar$ between the paths. 
Fig. \ref{fig:Ramsey_in_TJ} shows the Ramsey interference pattern from two Gaussian pulses as an example, the envelope function is $4P_1(\Delta\epsilon)$.

The probability to create a $e$-$h$ pair from a single voltage pulse is given by
\begin{equation} \label{P1}
 P_1\left(\Delta\epsilon\right)=\left\vert  \frac{T}{\hbar} \int\limits _{-\infty}^{\infty}e^{-i\varphi\left(t\right)}e^{-\frac{i}{\hbar}\Delta\epsilon t-0^+|t|}dt \right\vert^2
 \end{equation}
where $\varphi(t)=\frac{e}{\hbar}\int_{-\infty}^{t}{V_{0}(t')dt'}$. In order to relate \eqref{P2} to the Ramsey interference expression familiar for two-level systems, we consider rectangular voltage pulses of width $\tau$ and height $V_0$: $V(t)=V_0\left[\Theta(t)\Theta(\tau-t)+\Theta(t-t_0)\Theta(\tau-t-t_0)\right]$. 
In the limit of a small perturbation $V_0\ll\Delta\epsilon$, we check that the single pulse envelope corresponds to the two-level system  Rabi term \cite{scully1997}
\begin{equation} 
P_1\left(\Delta\epsilon,\tau\right)=\left\vert \frac{ T}{\Delta\epsilon}\right\vert^2\sin^2\left(\frac{\Delta\epsilon\tau/\hbar+\varphi_0}{2}\right).
\end{equation}

A voltage pulse excites a large number of $e$-$h$ pairs with different energies. The probability to create each pair depends on its energy $\Delta\epsilon$. To lowest order in $T$, there is no interference between pairs with different $k$ and $q$. A selection mechanism is needed  to isolate the probability $P_2(\Delta\epsilon)$ for the creation of  $e$-$h$ pairs with a given energy $\Delta\epsilon$.  We now show that the current noise dependence on $eV_{dc}$ provides such a mechanism. 

Provided that the temperature of the leads is much smaller than $eV_{dc}$, so we can benefit from the singularity of the Fermi distribution  $f'(\epsilon)\approx\delta(\epsilon)$,  it can be shown that the second derivative of the shot noise with respect to the $dc$ voltage  selects the value of the probability $P_2(\Delta\epsilon)$ at the specific energy $\Delta\epsilon = eV_{dc}$, namely \footnote{See Supplemental Material at [URL will be inserted by publisher] for the relation between the current noise and the creation probability.}
\begin{equation} \label{mainres1}
\frac{d^{2}S\left(\Omega\rightarrow0\right)}{d\left(eV_{dc}\right)^{2}}	
=e^2 \rho^2  P_2(\Delta\epsilon=eV_{dc})
\end{equation}
where $\rho$ is the density of states at the Fermi energy in the leads and the noise is defined as the Fourier transform of the current correlation function
$S(\Omega)=\int_{-\infty}^{\infty}dt_1\int_{-\infty}^{\infty}dt_2C(t_1,t_2)e^{i\Omega ( t_1 -  t_2 )}$, with $C(t_1,t_2)=\langle \hat{I}(t_1)\hat{I}(t_2)\rangle-\langle\hat{I}(t_1)\rangle\langle\hat{I}(t_2)\rangle$. To obtain Eq. \eqref{mainres1}, we have assumed that $\rho$ and $T$ are independent of energy. The peak of the single pulse envelope is proportional to the tunneling probability $|T|^2$, $P_1(\Delta\epsilon)\propto\frac{|T|^2}{\Delta\epsilon^2}$, a quantity which can be measured independently from a tunneling conductance measurement across the TJ.

Next we investigate the effect of interactions on the Ramsey interference pattern. We expect that dephasing processes will attenuate the interference term in \eqref{P2}, leaving the incoherent part unchanged.
As an example we consider the interaction between the electrons in the TJ and the electromagnetic environment modelled using $P(E)$ theory \cite{IngoldNazarov1992,Yurke1984,Devoret1997,FEYNMAN1963,CaldeiraLeggett1983},  by an  external electric circuit  of impedance $Z(\omega)$. 
This allows us to access Nyquist noise information and to probe charging effects with resistances small compared to the quantum or resistance.

The external impedance induces voltage fluctuations $\delta \hat{V}(t)$ on the tunnel junction, thereby modifying the energies of the electrons in the left lead by $\epsilon_{Lk} \rightarrow \epsilon_{Lk} +e\delta \hat{V}(t)$.
The fluctuating energy modifies the phase accumulated by an $e$-$h$ pair between the voltage pulses. Let us consider the tunnelling of an electron from the left to the right lead (see fig. \ref{fig:Ramsey_illustration}). At time $t_0$ (after the second pulse) the phase of the $e$-$h$ pair created in the first pulse is $\Delta\epsilon \,  t_0+\hat{\phi}(t=0)$ while the phase of the $e$-$h$ pair created in the second pulse is $\hat{\phi}(t=t_0)$, where $\hat{\phi}(t)\equiv\frac{e}{\hbar}\int_{-\infty}^{t}{\delta \hat{V}(t')dt'}$ is the phase operator of the TJ driven by the environment.
The effect of such voltage fluctuations could also describe the effect of electron-electron interactions \cite{Akkermans2007,cohen1989photons,Galaktionov2003,Kindermann2004}.

Using the $P(E)$ theory, we find that the dephasing due to the voltage fluctuations leads to an exponential decrease of the amplitude of the interference term \footnote{See Supplemental Material at [URL will be inserted by publisher] for the effect of the environment on the Ramsey interference pattern.}
\begin{equation} \label{P2wEnv}
P_{2}^{env}(\Delta\epsilon,t_0)\propto
2+2e^{J_R(t_0)}\cos\left(\frac{\Delta\epsilon \,  t_{0}}{\hbar} \pm\varphi_{0}\pm J_I(t_0) \right)
\end{equation}
where $J(t)\equiv J_R(t)+iJ_I(t)$ is a function of the environment only, and is defined as the correlation function of the TJ phase \cite{IngoldNazarov1992}
\begin{equation}
J(t)\equiv\langle\hat{\phi}(t)\hat{\phi}(0)\rangle-\langle\hat{\phi}^2(0)\rangle.
\end{equation}
Note that $J_R(t)<0$ for all $t$ so that the amplitude of the interference term is always attenuated. The environment is assumed to be, initially, at thermal equilibrium. Therefore, $\langle\cdot\rangle\equiv \text{Tr}\{\cdot e^{-\beta H_{env}}\}$ represents a thermal average over the environment degrees of freedom.
Note that we allow different temperatures for the environment and the system \cite{pekola2015,Jordan2015}. The environment (external impedance $Z(\omega)$) is at inverse temperature $\beta$ while the system (TJ) is at zero temperature (to satisfy eq. \eqref{mainres1}).

Eq. \eqref{P2wEnv} demonstrates the information that can be read off a Ramsey measurement:  the time-dependence of the environment correlations can be measured from the amplitude and the shift of the Ramsey fringes by varying $t_0$ the time between the voltage pulses. 

\begin{figure}
\begin{center}
\includegraphics[width=0.5\textwidth]{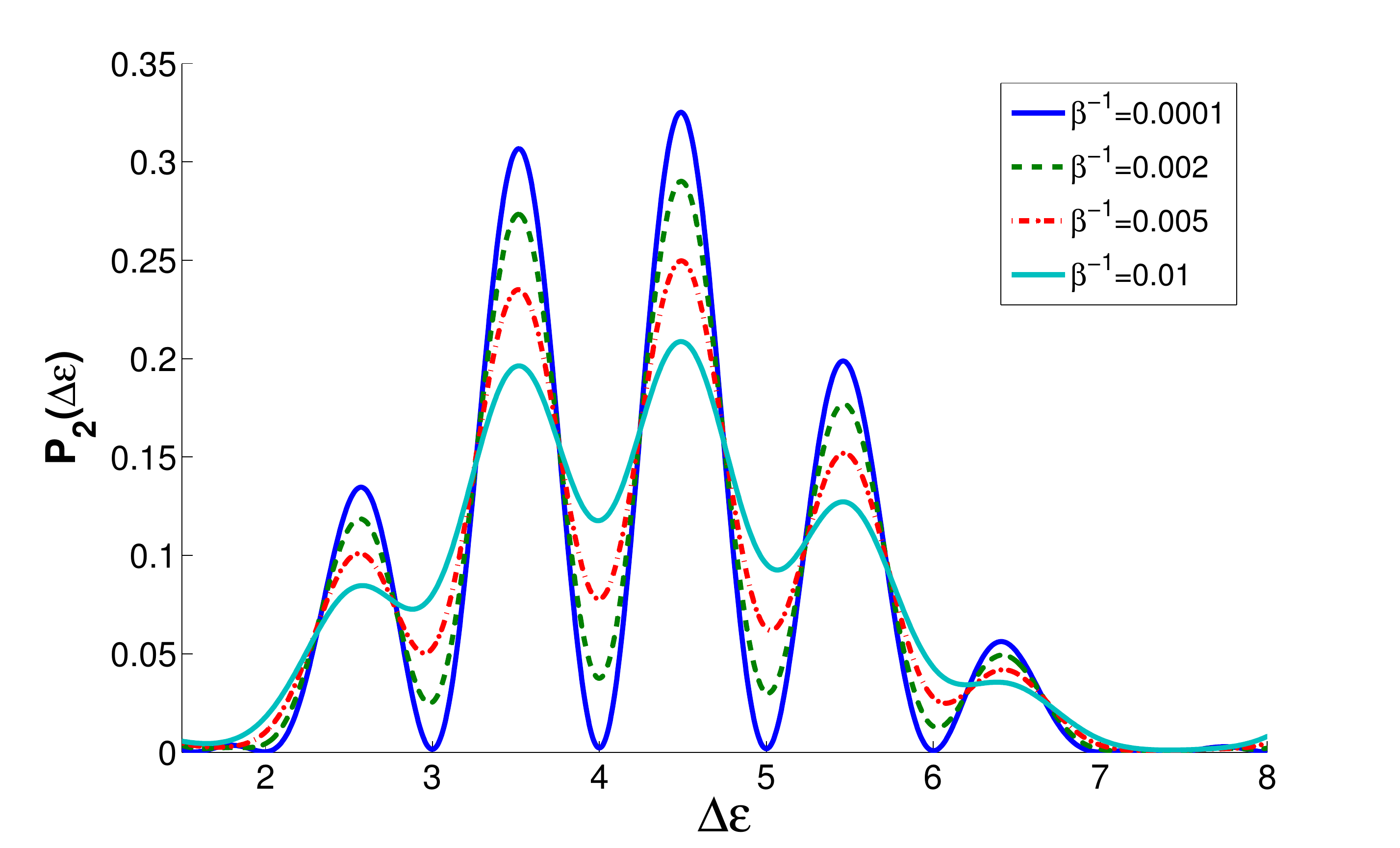}
\end{center}
\protect\caption[ Ramsey interference for the probability the creation of a $e$-$h$ pair in the Coulomb blockade regime for different temperatures]{ (color online) \it   Ramsey interference in the probability for  the creation of an $e$-$h$ pair  in the dynamical Coulomb blockade regime as a function of energy for different temperatures of the resistor but at fixed $t_0$ between the pulses.  The temperatures are given in units of the charging energy $E_c$. The voltage modulation and tunnelling matrix for all the plots  are  as in Fig. \ref{fig:Ramsey_in_TJ} and the energy $\Delta\epsilon$ is given in units of $\frac{2\pi\hbar}{t_0}$.  In order to  emphasize the $E_c$ shift of the interference pattern relative to Fig. \ref{fig:Ramsey_in_TJ}, we set the charging energy of the tunnel junction to be $E_c=1$ in units of $\frac{2\pi\hbar}{t_0}$.
\label{fig:bigR}
}
\end{figure}

For an ohmic environment $Z(\omega)=R$ the equation of motion of the TJ phase $\hat{\phi}(t)$ is analogous to that of a quantum Brownian particle with friction coefficient $(RC)^{-1}$ where $C$ is the tunnel junction capacitance \cite{IngoldNazarov1992}.
In the high temperature limit $\hbar\beta\ll RC$,  the limiting behavior of the correlation function $J_R(t)$ is \cite{QBM}
\begin{equation} \label{J_R}
J_R\left(t\right)=\begin{cases}
-\frac{1}{\hbar\beta}\frac{\pi}{R_{K}C}t^{2} & t\ll RC\\
-\frac{R}{R_{K}}\frac{2\pi}{\hbar\beta}\left(t-RC\right) & t\gg RC
\end{cases}
\end{equation}
where $R_K=h/e^2=25.8\text{k}\Omega$ is the quantum of resistance.
The correlation function $J(t)$ accounts for both the diffusive ($t\gg RC$) and ballistic ($t\ll RC$) evolution of the $TJ$ phase $\hat{\phi}(t)$. In both regimes the correlation $J_R(t)$ decreases linearly with temperature.
These limits are commonly associated with a high/low resistance limit.
The Ramsey method allows to investigate the ballistic-diffusive crossover of $\hat{\phi}(t)$ by tuning the time $t_0$ between the two voltage pulses while keeping the resistance $R$ constant and small compared to $R_K$.

In the diffusive regime the voltage fluctuations are described by Johnson-Nyquist noise. Taking $t_0\gg RC$ and assuming that the one-pulse envelope  varies slowly ($\frac{d}{d\Delta\epsilon}P_{1}^{0}\left(\Delta\epsilon\right)\ll \frac{\tau_d}{\hbar}$) the $e$-$h$ creation probability can be approximated by
\begin{equation} \label{P2_diffusion}
P_{2}^{env}\left(\Delta\epsilon,t_0\right)\approx2P_{1}^{0}\left(\Delta\epsilon\right)\left(1+e^{-\frac{t_0}{\tau_{d}}}\cos\left(\frac{\Delta\epsilon t_{0}}{\hbar}\pm\varphi_{0}\right)\right)
\end{equation}
where $\tau_{d}=\frac{\hbar\beta R_K}{2\pi R}$ is the typical dephasing time. In this regime the exponent is linear with $t_0$ and in the temperature $\beta^{-1}$ of the resistor.
This time $\tau_d$ was measured by a Mach-Zehnder interferometer \cite{Hansen2001} and has been predicted in Ref. \cite{Seelig2001}.

Interestingly enough, in the opposite ballistic regime ($t_0\ll RC$) where dynamical Coulomb blockade shows up, the Ramsey fringes are shifted by the charging energy $E_c=e^2/2C$ of the TJ. 
In the limit $t_0\ll RC$ and assuming that the single-pulse envelope varies slowly, ($\frac{d}{d\Delta\epsilon}P_{1}^{0}\left(\Delta\epsilon\right)\ll\frac{\tau_b}{\hbar} $), the $e$-$h$ creation probability can be approximated by
\begin{multline} \label{P2_Coloumb}
P_{2}^{env}\left(\Delta\epsilon,t_0\right)\approx2P_{1}^{0}\left(\Delta\epsilon+E_{c}\right) \\
\times\left(1+e^{-(\frac{t_{0}}{\tau_{b}})^2}\cos\left(\frac{\left(\Delta\epsilon\pm E_{c}\right)t_{0}}{\hbar}\pm \varphi_{0}\right)\right) 
\end{multline}
where $\tau_{b}=\hbar\sqrt{\frac{\beta}{E_c}}$ is the typical time for dephasing. In this regime the exponential decrease is quadratic in time but linear in the temperature $\beta^{-1}$ of the resistor. The shift  by $E_c$ implies that the tunnelling electron must compensate for the charging energy $E_c$ when hopping from left to right. Fig. \ref{fig:bigR} shows the decrease in the amplitude of the Ramsey fringes  with temperature
 in the dynamical Coulomb blockade regime \eqref{P2_Coloumb}. In all the plots, the time $t_0$ between the pulses is kept constant. 

The imaginary part  $J_I(t)$ of the correlation function does not depend on the environment temperature \cite{QBM},  and thus so does the Coulomb blockade shift for $t_0\gg RC$ (although the amplitude of the interference pattern does depend on it). We remind that the leads are  at zero temperature.
Using the Ramsey method, the dynamical Coulomb blockade shift can be observed  also for $R\ll R_K$, whereas in recent $dc$ transport experiments, the observation of a gap at low voltages requires $R\gg R_K$ \cite{jezouin2013}.

\begin{figure}
\begin{center}
\includegraphics[width=0.5\textwidth]{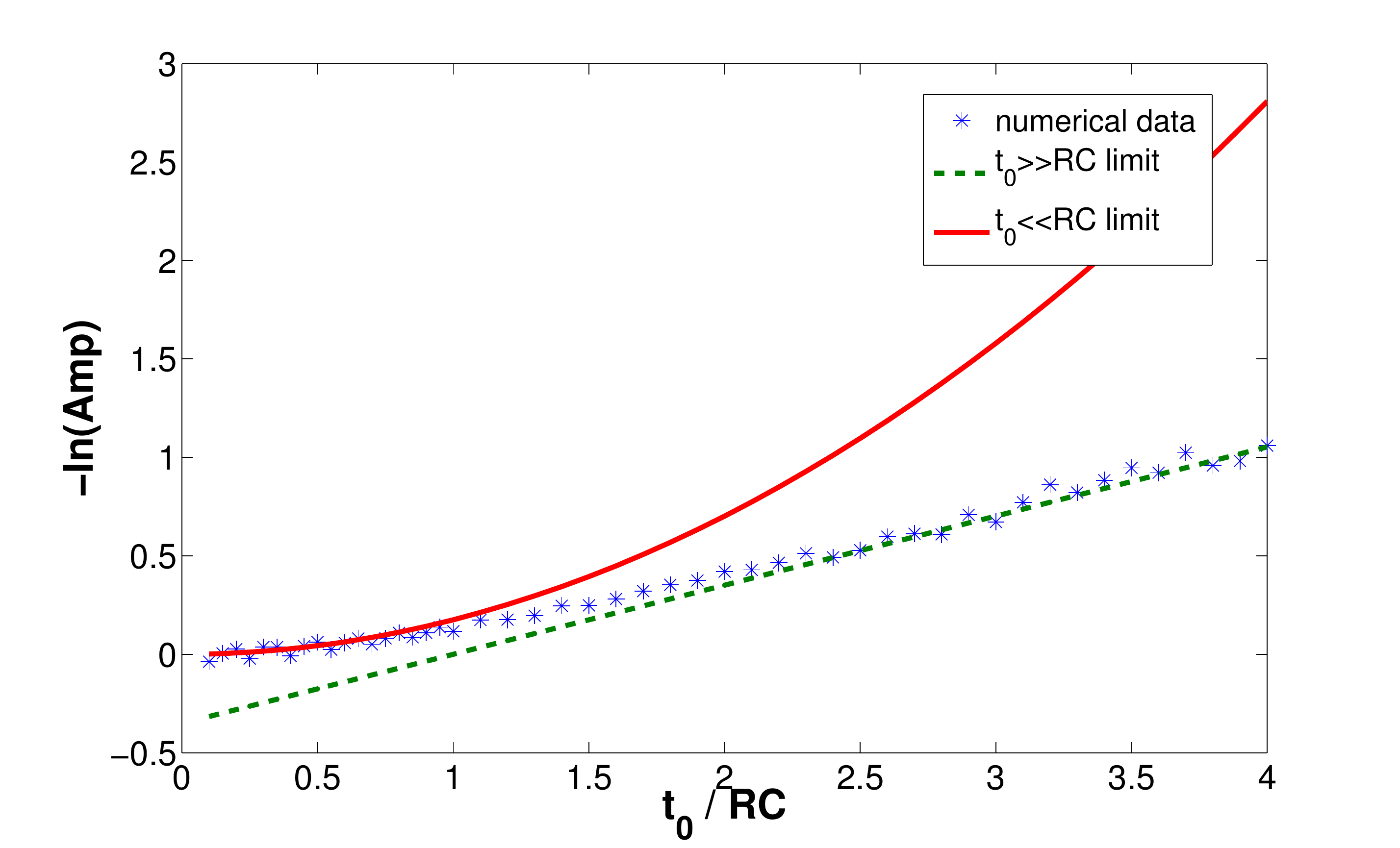}
\end{center}
\protect\caption[Crossover between ballistic and diffusive regimes]{ (color online) \it  The amplitude of the Ramsey fringes in the case of on ohmic environment as a function of the time between the voltage pulses $t_0$. The blue dots are numerical data the red and green lines are the approximations for small and large times. The system parameters are chosen to be $Ec=1.6\times10^{-4}\text{K}$, $\beta^{-1}=1.6\text{K}$, $\frac{R_K}{R}=750$. The voltage pulse is taken to be a Gaussian of width $\tau=5 \times 10^{-3} \, RC$ and pulse area $\varphi_0=7\pi$. 
\label{fig:Vis_vs_t0}
}
\end{figure}

To observe the crossover between the ballistic and diffusive regimes of the TJ phase $\hat{\phi}(t)$ in the high temperature limit ($\beta E_c\ll \pi\frac{R}{R_K}$) requires a precise fine-tuning of the parameters such that $t_0< ( \tau_d , \tau_b)$. Fig. \ref{fig:Vis_vs_t0}  displays this  ballistic to diffusive crossover around the time $t_0=RC$ between the two pulses as measured by the strength of the Ramsey fringes for the ohmic environment. The system parameters are chosen to be $Ec=1.6\times10^{-4} \text{K}$, $\beta^{-1}=1.6 \text{K}$, $R_K/R=750$.


By virtue of the uncertainty principle, the quantum bath is fluctuating even at very low temperature ($\hbar\beta\gg RC$). The long time limit of the correlation function is logarithmic, $J_R(t\gg RC)=-2\frac{R}{R_K}\left[\ln\left(\frac{t}{RC}+\gamma\right) \right]$, here $\gamma\approx 0.577$ is the Euler constant \cite{QBM}. This yields a power law attenuation of the prefactor of the Ramsey interference pattern
\begin{multline}\label{P2_0temp}
P_{2}^{env}(\Delta\epsilon,t_0)\propto 2+ 
2e^{-2\frac{R}{R_K}\gamma}\left(\frac{t_0}{RC}\right)^{-2\frac{R}{R_K}}\times \\
\cos\left(\Delta\epsilon t_{0}/\hbar \pm\varphi_{0}\pm \pi\frac{R}{R_K} \right)
\end{multline}
in contrast to the exponential behaviour in the high temperature limit of the bath.  This power law behavior gives rise to the zero bias anomaly of the conductance 
, and can be inferred from the analogy to the Luttinger liquid \cite{Safi2004,LeHur2005}.

To conclude, we have presented a manifestation of a phenomenon well-known in atomic optics, the Ramsey interference, in the realm of quantum mesoscopic system. 

Starting from Eq. \eqref{mainres1} for the zero frequency current noise as a function of $eV_{dc}$, we have presented the principles of a new spectroscopy of the excitations of a TJ, based on Ramsey interferometry which allows to explore interaction effects in the time domain. 
The Ramsey interference fringes can be used to measure the charging energy of a TJ and to study dynamical Coulomb blockade physics for resistances smaller than the quantum of resistance $R\ll R_K$. Moreover, the crossover between the "high/low" impedance behavior can be observed for a constant resistance by scanning the time $t_0$ between the voltage pulses. We have shown that Ramsey interference is a powerful tool to study the time dependence of dephasing processes.
We anticipate that this protocol could also lead to interesting and rich interferometry patterns in Luttinger liquids \cite{Safi2004,LeHur2005}, quantum Hall and other topological systems \cite{atala2013}.

Acknowledgements:  This work was supported by the Israel Science Foundation Grant No.924/09, by the Labex PALM Paris-Saclay ANR-10-LABX-0039 and by the DFG Forschegruppe 2014. We thank Julien Gabelli for discussions. This work has also beneffited from discussions at CIFAR meetings. 

\bibliographystyle{apsrev4-1}
%



\section*{Supplementary material}

# Ramsey Interferometry of Particle-Hole Pairs in Tunnel Junctions

Tal Goren[1,2], Karyn Le Hur[2] and Eric Akkermans[1]

[1]*Department of Physics, Technion Israel Institute of Technology, Haifa 32000, Israel and*
[2]*Centre de Physique Théorique, École Polytechnique, CNRS,*
*Université Paris-Saclay, 91128 Palaiseau Cédex, France*
(Dated: November 15, 2016)

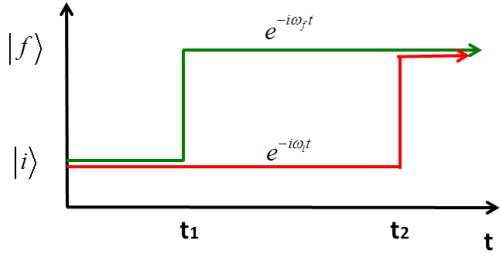

FIG. 1. *Schematic representation of the two paths of the system to go from the initial state $|i\rangle$ to the final state $|f\rangle$ and the phases the system accumulates.*

## RAMSEY INTERFERENCE

The aim of this section is first to recall the Ramsey interference protocol in the context of a two-level atom. Then, we consider the generalization to a continuum set of energy levels, applied to the tunnel junction.

### Two level atom

Consider a two level atom with states $|i\rangle$ and $|f\rangle$ and energies $E_i$ and $E_f$, respectively. Suppose that a transition between the states, with transition amplitude $A_{if}^{(1)}$, can occur only at two times, $t = 0$ and $t = t_0$.

If the atom was initially in state $|i\rangle$, two distinct paths can be considered to reach the state $|f\rangle$ (see Fig. 1). In the first path the atom stays in state $|i\rangle$ at $t = 0$ and transfers to state $|f\rangle$ at $t = t_0$, in the second path it reaches state $|f\rangle$ at time $t = 0$ and remains there. Since the states have different energies, the phase accumulated along the two paths is different. If the transition probabilities are small $P_{if}^{(1)} = |A_{if}^{(1)}|^2 \ll 1$, such that the transition probability at $t_2$ $P_{if}^{(2)} \propto P_{if}^{(1)}(1 - P_{if}^{(1)}) \approx P_{if}^{(1)}$, the amplitudes for the different paths can be approximated as

$$A_1(t_0) = e^{\frac{-iE_i t_0}{\hbar}} A_{if}^{(1)} \tag{1}$$

$$A_2(t_0) = e^{\frac{-iE_f t_0}{\hbar}} A_{if}^{(1)}. \tag{2}$$

The transition amplitude is a sum of all the possible paths, and the corresponding probability for the tran-

sition is

$$P_{if}^{(2)} = |A_1 + A_2|^2 = 4\cos^2\frac{(E_f - E_i)t_0}{2\hbar} P_{if}^{(1)}. \tag{3}$$

Hence, the Ramsey setup is a two slit experiment where the phase is accumulated over time instead of space [1]. It can be used to obtain the energy difference between the states $|i\rangle$ and $|f\rangle$, *i.e.* to perform a spectroscopy of the quantum system.

By applying two consequent modulations on a quantum system, the transition probability between one state of the system to another can be enhanced or reduced depending on the time difference between the consequent modulations.

Next we describe the generalization of Ramsey effect to particle-hole pairs in a mesoscopic system, the tunnel junction.

### Tunnel junction

A tunnel junction is composed of two metallic parts separated by a thin insulating barrier. The barrier is thin enough to allow charge transport by tunneling of electrons. The tunneling of electrons is induced by applying a voltage on the tunnel junction. When an electron tunnels, e.g. from the right lead to the left lead, it leaves a hole in the right lead. Therefore, a tunneling event creates an e-h pair in the tunnel junction. Note, that the electron and hole are in different leads.

By applying two consequent voltage pulses separated by $t_0$ on the tunnel junction the probability to create an e-h will exhibit Ramsey interference. The initial state of the leads is the Fermi sea at zero temperature, i.e. the vacuum state for electrons and holes

$$|i\rangle = \prod_{k,q} |0_k, 0_q\rangle \tag{4}$$

where $k$ marks the momentum of the electrons and $q$ the momentum of the holes. A voltage pulse creates a large number of e-h pairs, each with a probability amplitude $A_{kq}$. The system evolves freely for a time $t_0$, during which the e-h states accumulate a phase $(\epsilon_k - \epsilon_q) t_0/\hbar$. When the second voltage pulse is applied, the same set of e-h pairs can be created from the vacuum part of the wave function. Taking only the lowest order processes in $A_{kq}$,

the time evolution can be summarized as follows

$$
|i\rangle \xrightarrow{\text{first pulse}} |i\rangle + \sum_{k,q} A_{kq}|1_k 1_q\rangle
$$

$$
\xrightarrow{\text{time evolution}} |i\rangle + \sum_{k,q} A_{kq} e^{-i(\epsilon_k-\epsilon_q)t/\hbar}|1_k 1_q\rangle \qquad (5)
$$

$$
\xrightarrow{\text{second pulse}} |i\rangle + \sum_{k,q} A_{kq}\left(1 + e^{-i(\epsilon_k-\epsilon_q)t_0/\hbar}\right)|1_k 1_q\rangle
$$

Therefore, the probability to create an e-h pair after two voltage pulses is given by

$$
P_{kq}^{(2)} = 4\cos^2\frac{(\epsilon_k-\epsilon_q)t_0}{2\hbar} P_{kq}^{(1)} \qquad (6)
$$

just as in the simple case of a two level atom.

Each e-h pair is analogous to one two level atom, the state without a pair correspond to the ground state and the creation of a pair corresponds to a transition into the exited state. Therefore, a tunnel junction can be thought of as a collection of infinite number of two level atoms with different energy gaps. Each of these atoms can be excited by the voltage pulses and exhibits Ramsey interference.

A selective mechanism is needed in order to observe the probability $P(\Delta\epsilon)$ for the creation of e-h pairs with a given energy $\Delta\epsilon$. We show that the current noise dependence on $eV_{dc}$ provides a measurement of the probability $P(\Delta\epsilon)$. Therefore, we provide a way to observe Ramsey interferences in mesoscopic physics.

## RELATION BETWEEN THE NOISE AND THE CREATION PROBABILITY

The electrons in the two metal electrodes are described by the Hamiltonian

$$
H_0 = \sum_k \epsilon_{Lk}(t) c_{Lk}^\dagger c_{Lk} + \sum_q \epsilon_{Rq} c_{Rq}^\dagger c_{Rq} \qquad (7)
$$

where $c_{Lk}^\dagger$ $\left(c_{Rq}^\dagger\right)$ creates an electron in the left (right) lead with momentum $k$ $(q)$. We choose the gauge

$$
\epsilon_{Lk}(t) = \epsilon_{Lk} - eV_{dc} + eV(t) \equiv \epsilon_{Lk}^0 + eV(t)
$$
$$
\epsilon_{Rq} = \epsilon_{Rq} \qquad (8)
$$

although our results do not depend on the gauge.

Tunneling is introduced by the Hamiltonian

$$
H_T = \sum_{kq}\left[ T_{kq} c_{Rq}^\dagger c_{Lk} + T_{kq}^* c_{Lk}^\dagger c_{Rq} \right] \qquad (9)
$$

such that $T_{kq} c_{Rq}^\dagger c_{Lk}$ transfers an electron from the left lead to the right lead. The probability of a tunneling event is small, i.e. the tunneling matrix elements are small compared to the Fermi energy ($|T_{kq}| \ll \epsilon_F$).

A tunneling event from the initial state creates an e-h pair with total energy $\Delta\epsilon = \pm(\epsilon_{Lk}^0 - \epsilon_{Rq})$, the $\pm$ sign accounts for tunnelling processes from the left/right lead to the right/left lead. We consider the zero temperature limit for the leads, therefore the initial state of the system is the vacuum sate of the electrons and holes $|00\rangle$. The e-h states correspond to

$$
|1_{e_{Lk}} 1_{h_{R,-q}}\rangle = c_{Lk}^\dagger c_{Rq}|i\rangle
$$
$$
|1_{e_{Rq}} 1_{h_{L,-k}}\rangle = c_{Rq}^\dagger c_{Lk}|i\rangle \qquad (10)
$$

## Probability to create an electron-hole pair

The total probability to create an e-h pair to second order in $T_{kq}$ is

$$
P_{tot} = \sum_{\substack{k>k_F\\q<k_F}} P_{LR}^{k,-q} + \sum_{\substack{k<k_F\\q>k_F}} P_{RL}^{q,-k} \qquad (11)
$$

where $P_{LR}^{k,-q}$ is the probability to create an electron with momentum $k$ in the left lead and a hole with momentum $-q$ in the right lead from the vacuum, and $P_{RL}^{q,-k}$ is the probability to create an electron in the right lead with momentum $q$ and a hole in the left lead with momentum $-k$. To second order in $T_{kq}$ these processes do not interfere.

$$
P_{LR}^{k,-q} = |\langle 1_{e_{Lk}} 1_{h_{R,-q}}|U(\infty,-\infty)|0_{e_{Lk}} 0_{h_{R,-q}}\rangle|^2
$$
$$
P_{RL}^{q,-k} = |\langle 1_{e_{Rq}} 1_{h_{L,-k}}|U(\infty,-\infty)|0_{e_{Rq}} 0_{h_{L,-k}}\rangle|^2 \qquad (12)
$$

where $U(t_f, t_i)$ is the evolution operator which satisfies $i\frac{d}{dt}U(t) = H(t)U(t)$ and the $|00\rangle$ state correspond to the Fermi sea in both leads (the initial state of the system).

The probability to create an e-h pair to second order in $T_{kq}$ can be expressed as

$$
P_{LR}^{k,-q} = \frac{1}{\hbar^2}\int_{-\infty}^{\infty} dt_1 dt_2 \langle 0_{e_{Lk}} 0_{h_{R,-q}}|\tilde{H}_T(t_1)|1_{e_{Lk}} 1_{h_{R,-q}}\rangle
$$
$$
\times \langle 1_{e_{Lk}} 1_{h_{R,-q}}|\tilde{H}_T(t_2)|0_{e_{Lk}} 0_{h_{R,-q}}\rangle \qquad (13)
$$
$$
= \left|\langle i|c_{Rq}^\dagger c_{Lk} c_{Lk}^\dagger c_{Rq}|i\rangle\right|^2
$$
$$
\times \left|\frac{T_{kq}}{\hbar}\int_{-\infty}^{\infty} e^{-i\varphi(t)} e^{-\frac{i}{\hbar}(\epsilon_{Lk}^0-\epsilon_{Rq})t} dt\right|^2
$$
$$
= \Theta(\epsilon_{Lk}-\epsilon_F)\left(1 - \Theta(\epsilon_{Rq}-\epsilon_F)\right)
$$
$$
\times \left|\frac{T_{kq}}{\hbar}\int_{-\infty}^{\infty} e^{-i\varphi(t)} e^{-\frac{i}{\hbar}(\epsilon_{Lk}^0-\epsilon_{Rq})t} dt\right|^2 \qquad (14)
$$

where we define $\varphi(t) \equiv \frac{e}{\hbar}\int_{-\infty}^t V(t') dt'$ and $\Theta(\epsilon)$ is the heaviside function. The time evolution of $\tilde{H}_T$ is with

respect to the bare Hamiltonian $H_0$, i.e. $\tilde{H}_T(t) = U_0^\dagger(t) H_T U_0$ where $U_0(t) = e^{-\frac{i}{\hbar}\int_{-\infty}^t H_0(t')dt'}$.

In the same way

$$P_{RL}^{q,-k} = \Theta(\epsilon_{Rq} - \epsilon_F)\left(1 - \Theta(\epsilon_{Lk} - \epsilon_F)\right)$$
$$\times \left| \frac{T_{kq}}{\hbar} \int_{-\infty}^{\infty} e^{+i\varphi(t)} e^{-\frac{i}{\hbar}\left(\epsilon_{Rq} - \epsilon_{Lk}^0\right)t} dt \right|^2. \quad (15)$$

The last line in eq. (13) states that the probability to create an e-h pair is just the Fourier transform of $g(t) = e^{-i\varphi(t)}$ at frequency $\epsilon_{Lk}^0 - \epsilon_{Rq}/\hbar$. Note that the pair-creation probability depends only on the energy difference $\Delta\epsilon = \pm(\epsilon_{Lk}^0 - \epsilon_{Rq})$.

The tunneling Hamiltonian does not connect $|00\rangle$ with itself, therefore $|11\rangle\langle 11|$ in eq. (13) can be replaced by the unit operator to obtain the total probability to create an e-h pair

$$P_{total} = \frac{1}{\hbar^2} \int_{-\infty}^{\infty} dt_1 dt_2 \langle i| \tilde{H}_T(t_1) \, \tilde{H}_T(t_2) |i\rangle \quad (16)$$

**Current noise**

The current through the conductor is defined as the change in time of the number of electrons in the left lead.

$$I(t) = e\frac{d}{dt}\sum_k c_{Lk}^\dagger c_{Lk} = -i\frac{e}{\hbar}\sum_k \left[ c_{Lk}^\dagger c_{Lk}, H \right]$$
$$= i\frac{e}{\hbar}\sum_{kq} \left[ T_{kq} c_{Rq}^\dagger c_{Lk} - T_{kq}^* c_{Lk}^\dagger c_{Rq} \right]. \quad (17)$$

The power spectrum (the noise) is defined as the Fourier transform of the current correlation function

$$S(\Omega) = \int_{-\infty}^{\infty} dt_1 \int_{-\infty}^{\infty} dt_2 C(t_1, t_2) e^{i\Omega t_1} e^{-i\Omega t_2}$$
$$C(t_1, t_2) = \langle \hat{I}(t_1)\hat{I}(t_2)\rangle - \langle \hat{I}(t_1)\rangle\langle \hat{I}(t_2)\rangle \quad (18)$$

where $\hat{I}(t)$ is the current operator in the Heisenberg picture. Since the current operator is linear in $T$, the correlation function to second order in $T$ is given by

$$C(t_1, t_2) = \langle \tilde{I}(t_1)\hat{I}(t_2)\rangle - \langle \tilde{I}(t_1)\rangle\langle \hat{I}(t_2)\rangle \quad (19)$$

where the time evolution of $\tilde{I}(t)$ is with respect to the bare Hamiltonian $H_0$.

The tunneling Hamiltonian and the current operator act on the $|00\rangle$ state in the same way up to a factor, therefore we can express the total probability to create an e-h pair as

$$P_{total} = \frac{1}{e^2} \int_{-\infty}^{\infty} dt_1 dt_2 \langle i| \tilde{I}(t_1) \, \tilde{I}(t_2) |i\rangle \quad (20)$$

which is just the current noise at zero frequency.

$$S(\Omega \to 0) = e^2 \left[ \sum_{\substack{k > k_F \\ q < k_F}} P_{LR}^{k,-q} + \sum_{\substack{k < k_F \\ q > k_F}} P_{RL}^{q,-k} \right]. \quad (21)$$

Taking the continuum limit and assuming that the density of states and the tunneling matrix elements do not depend on energy we can write the noise as

$$S(\Omega \to 0) = e^2 \rho^2 \left[ \int_{\epsilon_F}^{\infty} d\epsilon_{Lk} \int_{-\infty}^{\epsilon_F} d\epsilon_{Rq} P_{LR}(\Delta\epsilon = \epsilon_{Lk} + eV_{dc} - \epsilon_{Rq}) + \int_{-\infty}^{\epsilon_F} d\epsilon_{Lk} \int_{\epsilon_F}^{\infty} d\epsilon_{Rq} P_{RL}(\Delta\epsilon_{Rq} - \epsilon_{Lk} - eV_{dc}) \right] \quad (22)$$

where $\rho$ is the density of states of the leads at the Fermi energy. Note that changing the integration variable $\epsilon_{Lk}$ to $\epsilon_{Lk} + eV_{dc}$ is equivalent to having a difference of $eV_{dc}$ in the chemical potentials of the electrodes.

Taking the second derivative of the noise with respect to the dc voltage we get eq. (7) in the article

$$\frac{d^2 S(\Omega \to 0)}{d(eV_{dc})^2} = e^2 \rho^2$$
$$\times \left[ P_{RL}^{q,-k}(\Delta\epsilon = eV_{dc}) + P_{LR}^{k,-q}(\Delta\epsilon = -eV_{dc}) \right]. \quad (23)$$

The second derivative of the shot noise with respect to

the dc voltage is proportional to the probability to create an e-h pair with energy $\pm V_{dc}$. Here we have identified a way to measure the probability to create an electron hole pair with specific energy in a tunnel junction. Note that for any voltage only one of these probabilities is non zero. The symmetry between left and right is indicated by $P_{RL}^{q,-k}(\Delta\epsilon) = P_{LR}^{k,-q}(-\Delta\epsilon)$

## EFFECT OF AN EM ENVIRONMENT ON THE RAMSEY FRINGES

The probability for the creation of e-h pairs in the presence of an environment is a convolution-like expression between the pair creation probability in the absence of environment and the environment absorption probability. To be more precise, for the Ramsey voltage modulation, the probability for the creation of a e-h pair after two pulses is given by

$$P_{LR,2}^{env}\left(\Delta\epsilon\right) = \int_{-\infty}^{\infty} dE P(E) P_{RL,1}^0(\Delta\epsilon + E) \times \left[2 + 2\cos\left(\frac{(\Delta\epsilon + E)t_0}{\hbar} + \varphi_0\right)\right] \quad (24)$$

where $P_{RL,1}^0$ is the pair creation probability due to a single pulse in the absence of the environment and $P(E)$ is the probability of the environment to absorb energy $E$ [2]. The convolution-like form of the probability suggests that the environment will suppress the Ramsey fringes. If the width of $P(E)$ is larger than the the period $\hbar t_0^{-1}$ of the Ramsey fringes the interference pattern will disappear.

To better understand the effect of the environment on the Ramsey interference let us consider the probability to create a e-h pair expressed in the time domain

$$P_{LR}^{env}\left(\Delta\epsilon_{LR}\right) = \frac{T^2}{\hbar^2} \int_{t_i}^{t_f} dt_1 \int_{t_i}^{t_f} dt_2 e^{-i\varphi(t_1) + i\varphi(t_2)} \times e^{-i\frac{\Delta\epsilon_{LR}}{\hbar}t_1} e^{+i\frac{\Delta\epsilon_{LR}}{\hbar}t_2} e^{J(t_2 - t_1)}. \quad (25)$$

Using the stationary phase approximation, the main contributions to the integral come from the times where the argument of the exponent is stationary. Assuming that $\dot{J} \ll \dot{\varphi}$ and $\dot{J} \ll (\epsilon_{Rq} - \epsilon_{Lk})$, the stationary condition is

$$\hbar\dot{\varphi}(t_1) = \Delta\epsilon_{LR} \quad (26)$$

$$\hbar\dot{\varphi}(t_2) = \Delta\epsilon_{LR}. \quad (27)$$

Remember that we consider only cases where $|\Delta\epsilon_{LR}| > 0$, therefore the main contribution to the integral comes from times satisfying

$$V(t_1) = V(t_2) \neq 0. \quad (28)$$

Assuming that the two pulses are narrow, i.e. $V(t) \approx \varphi_0\delta(t) + \varphi_0\delta(t - t_0)$, there are only 4 possibilities to satisfy the above condition:

$$\begin{aligned} t_1 &= t_2 = 0 \\ t_1 &= t_2 = t_0 \\ t_1 &= 0 \quad t_2 = t_0 \\ t_1 &= t_0 \quad t_2 = 0 \end{aligned} \quad (29)$$

and the e-h creation probability is approximately given by

$$P_{LR}^{env}\left(\Delta\epsilon_{LR}\right) \sim 2e^{J(0)} + e^{J(t_0)}e^{+i\varphi_0}e^{+i\frac{\Delta\epsilon_{LR}t_0}{\hbar}} + e^{J(-t_0)}e^{-i\varphi_0}e^{-i\frac{\Delta\epsilon_{LR}t_0}{\hbar}}$$

$$= 2 + 2e^{J_R(t_0)}\cos\left(\frac{\Delta\epsilon_{LR}t_0}{\hbar} + \varphi_0 + J_I(t_0)\right) \quad (30)$$

where we used the time reversal symmetry for the free environment $J(t) = J^*(-t)$ and we defined the real and imaginary part of the correlation function $J(t) = J_R(t) + iJ_I(t)$. This result can be understood in our two-slit description of Ramsey interferometry. The environment dephases the phase of the electron, therefore it reduces only the interference (quantum) part of the probability and keeps the classical sum of the two paths unchanged. Moreover, the amplitude of the Ramsey fringes decays exponentially with the correlation function of the environment $J(t_0)$. Therefore, by measuring it as a function of the time between the pulses, $t_0$, we can obtain the correlation function of the bath as a function of time.

The dephasing time is the characteristic time for the decay of the interference effect. For an Ohmic environment it is $\tau_b$ for the ballistic regime and $\tau_d$ for the diffusive regime, defined in the main text. Using Ramsey interferometry, the dephasing time can be measured in a direct way: the time $t_0$ between the pulses is increased until the Ramsey interference vanishes.

## TUNNEL JUNCTION PHASE AS A QUANTUM BROWNIAN PARTICLE

In order to find the correlation function $J(t)$ we need to solve the equation of motion for the phase $\hat{\phi}$ of the tunnel junction (without the coupling to the electron system). The Hamiltonian of the RC circuit is given by [2]

$$H_{bath} = \frac{\hat{Q}^2}{2C} + \sum_n\left(\frac{\hat{q}_n^2}{2C_n} + \left(\frac{\hbar}{e}\right)^2\frac{1}{2L_n}\left(\hat{\phi} - \hat{\phi}_n\right)^2\right) \quad (31)$$

where $\hat{Q}$ and $\hat{\phi}$ are the charge and phase of the tunnel junction, treated here as a capacitor. The resistor is modeled as an infinite sum of conjugated LC circuits. This is exactly the Hamiltonian describing a quantum brownian particle [3] where the particle quantities are

$$\begin{aligned} Q &\leftrightarrow p \\ \phi &\leftrightarrow x \\ C &\leftrightarrow M \end{aligned} \quad (32)$$

and the bath degrees of freedom are

$$q_n \leftrightarrow p_n \tag{33}$$
$$\phi_n \leftrightarrow x_n$$
$$C_n \leftrightarrow m_n$$
$$\frac{1}{\sqrt{L_n C_n}} \leftrightarrow \omega_n$$

From the classical equation of motion for the phase $\phi$ in an RC circuit is $\ddot{\phi} + \frac{1}{RC}\dot{\phi}(s) = 0$ we learn that the friction coefficient of the brownian particle is

$$\frac{1}{RC} \leftrightarrow \gamma. \tag{34}$$

The correlation function for a quantum brownian particle are known $J(t) = J_R(t) + iJ_I(t)$ [3]

$$J_I(t) = -\pi \frac{R}{R_K}\left\{1 - e^{-\frac{t}{RC}}\right\} \tag{35}$$

$$J_R(t) = \frac{e^2}{\hbar^2}\left[ -\frac{R}{\beta}t + \frac{R^2 C}{\beta} + \frac{2}{\beta R C^2}\sum_{n=1}^{\infty}\frac{e^{-\nu_n t}}{\nu_n\left(\frac{1}{R^2 C^2} - \nu_n^2\right)}\right]$$
$$- \frac{\hbar R}{2}\cot\left(\frac{\hbar\beta}{2RC}\right)e^{-\frac{t}{RC}} - \frac{2}{C\beta}\sum_{n=1}^{\infty}\frac{1}{\nu_n^2 + \nu_n\frac{1}{RC}}$$

where $\nu_n = \frac{2\pi n}{\hbar\beta}$.

The imaginary part of the correlation is independent on temperature and its limit for short and long times are

$$J_I(t) = \begin{cases} -\pi\frac{t}{R_K C} = -\frac{1}{\hbar}E_c t & t \ll RC \\ -\pi\frac{R}{R_K} & t \gg RC \end{cases} \tag{36}$$

The imaginary part of the correlation function is manifested by a shift in the Ramsey interference pattern (see eq. (30) above). The dynamical Coulomb blockade shift (eq. (12) in the letter) is observed for short times $t_0 \ll RC$. For long times $t_0 \gg RC$ the shift can be observed in the low temperature limit (eq. (13) in the letter) and is negligible in the high-temperature limit (eq. (11) in the letter)

Taking the high temperature limit $\hbar\beta \ll RC$ we get the limiting behavior of the correlation function $J_R(t)$

$$J_R(t) = \begin{cases} -\frac{1}{\hbar\beta}\frac{\pi}{R_K C}t^2 & t \ll RC \\ -\frac{R}{R_K}\frac{2\pi}{\hbar\beta}(t - RC) & t \gg RC \end{cases} \tag{37}$$

In the zero temperature limit $\hbar\beta \gg RC$ the limiting behavior of the correlation function $J_R(t)$ is

$$J_R(t) = \begin{cases} -2\frac{R}{R_K}\frac{1}{4}\left[3 - 2\gamma - 2\ln\left(\frac{t}{RC}\right)\right]\left(\frac{t}{RC}\right)^2 & t \ll RC \\ -2\frac{R}{R_K}\left[\ln\left(\frac{t}{RC}\right) + \gamma\right] & t \gg RC \end{cases} \tag{38}$$

where $\gamma \approx 0.577$ is the Euler gamma.



\end{document}